# IEEE Copyright Notice

Accepted to be published in:

- Proceedings of the 2020 - IEEE International Conference on Advanced Networks and Telecommunications Systems - Workshop on New Advances on Vehicle-to-Everything (V2X) Communications and Networking), 14-17 December 2020.

# A Survey on Blockchain and Edge Computing applied to the Internet of Vehicles

Anderson Queiroz, Eduardo Oliveira, Maria Barbosa and Kelvin Dias
*Centro de informática*
*Universidade Federal de Pernambuco*
Recife, Brazil
aalq@cin.ufpe.br, ehammo@cin.ufpe.br, mksb@cin.ufpe.br, kld@cin.ufpe.br

*Abstract*—With the advent of Intelligent Transportation Systems (ITS), data from various sensors embedded into vehicles or smart cities infrastructure are of utmost importance. This ecosystem will require processing power and efficient trust mechanisms for information exchange in vehicle-to-everything (V2X) communications. To accomplish these requirements, both edge computing and blockchain have been recently adopted towards a secure, distributed, and computation empowered Internet of Vehicles (IoV). This paper surveys prominent solutions for blockchain-based vehicular edge computing (VEC), provides a taxonomy, highlights their main features, advantages, and limitations to provide subsidies for further proposals.

*Index Terms*—VANET, Internet of Vehicles (IoV), Fog, Edge and Cloud Computing, Blockchain

## I. INTRODUCTION

Vehicular Ad-hoc Networks (VANETs) have emerged as a key technology to support road safety, comfort, and infotainment applications for passengers and citizens of smart cities. VANETs rely on the following communication modes: Vehicle-to-Infrastructure (V2I), Vehicle-to-Vehicle (V2V), and Vehicle-to-Everything (V2X). These networks aim to provide communication with any component of the Intelligent Transportation System (ITS) in smart cities, such as other vehicles, stations, pedestrians, traffic lights, bicycles, and any other member participating in this ecosystem. VANETs are composed of entities, such as sensors, cameras, and On-Board Computerized Units (OBU) located in vehicles, Road Side Units (RSU), and Base Stations (BS) with various communication network standards, such as: 802.11p, 802.15, 4G/LTE, and 5G/NR, which have more computational power than vehicle OBUs [1].

The data can be processed and delivered by the base station to the cloud or by vehicles collaboratively. Figure 1 presents the various architectural, technological, and communication aspects that embrace the IoV environment in the context of this article. However, to be widely implemented, VANETs still lack effective security and privacy mechanisms to minimize false and malicious exchanges of information between vehicles. Such incidents can cause different types of accidents that threaten the lives of drivers, passengers, and pedestrians [2]. Considering that even with the growth in the number of devices carrying out transactions on VANETs, it is possible to reach security levels through the features (immutability, cryptography, shared registration, distributed processing, and joint validation of data and transactions) offered by the blockchain.

Traditional access to the remote cloud may degrade VANETs services due to incurred latency. Nevertheless, with the emergence of Vehicular Edge Computing (VEC), processing and storage are now close to vehicles, that is, at the edge of the network. Recently, researchers proposed solutions for joint blockchain and edge computing applied to the Internet of Vehicles (IoV). Besides cloud-like services closer to the end-user (Edge, Fog [1], and Cloudlet), IoV also comprises softwarization technologies, such as Software-Defined Networking (SDN) and Network Functions Virtualization (NFV), as well, artificial intelligence support for decision making from both autonomous cars and automating service orchestration and management of vehicular infrastructure.

The proposed solutions have points in common and peculiarities in terms of architectural design and also distinct characteristics that must be clarified to understand their advantages and disadvantages in the scenario of future developments of possible solutions in the area. Various approaches involving different edge computing techniques and applications, blockchain, and other security mechanisms for exchanging information and data, have been introduced to the academic community and companies in recent years. This study aims to analyze and compare these different researches by correlating the proposals, intending to understand each technology, and how they work and deal with the environment in each application [3]–[8].

This article examines crucial solutions for edge computing and blockchain applied to IoV, provides a taxonomy, and highlights, through a comparative summary, its main proposals, technologies, and architectures as a way of providing subsidies to other proposals. Section II shows the background on Blockchain, Section III presents the taxonomy of the comparative elements of the study, the summary of the solutions, and the discussion of the advantages and limitations of each proposal. Section IV presents the conclusion of the article.

---

[1]This article considers the term Fog to refer to the end nodes, that is, the vehicles

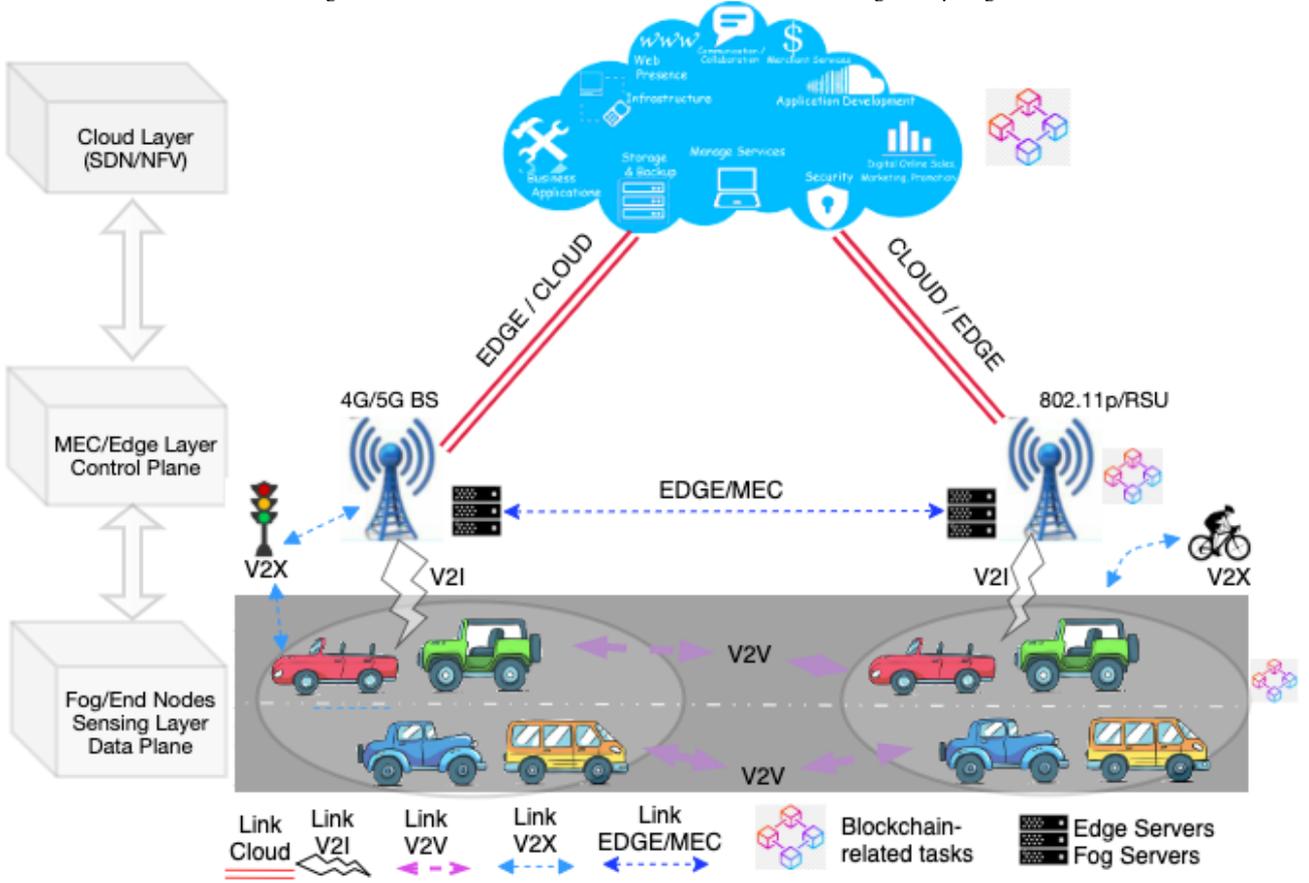

Fig. 1. A General Framework for Blockchain-based Vehicular Edge Computing

## II. FOUNDATION

### A. Blockchain

As Jianbin Gao said in [1] "Blockchain is a decentralized infrastructure and a distributed computing paradigm that uses encrypted chained block structures to validate and store data, consensus algorithms to generate and update data and smart contracts to program and manipulate data". These blocks are made up of transactions that are verified and validated in a distributed manner, that is, without the need for a central trust entity. This solution was first proposed by Satoshi Nakamoto [2] to define a way to validate transactions carried out on the internet with the bitcoin cryptocurrency. By its nature, Blockchain management is achieved by several participants, unlike traditional databases, which are generally managed by a single organization. Another important feature of the blockchain is its ability to easily track and verify data [1].

Currently, there are three ways to implement blockchain on a network: public, private, or consortium. The most common implementation used by cryptocurrencies is the public blockchain, where any node can participate in some activity of the management process. Private implementation is characterized by the centralized control of the nodes and rules by a single organization. Finally, the consortium blockchain, where a certain group of companies, usually with the same objective, perform the control and management of the nodes and their rules. This last type of implementation, consortium or federated, has been widely used in the IoV environment since its characteristics are similar to the current model of vehicular networks as demonstrated in table 1. [3].

We also need to keep in mind two blockchain concepts, The consensus algorithm and the smart contracts. The consensus algorithm aims to ensure higher trust and security on the network. Therefore, it verifies the authenticity of the transactions and validates the blocks. However, for this to happen, there must be a consensus on the network. In other words, the consensus algorithm is the means to do agreements among the network nodes, ensuring that the terms designated by the protocol will be met. The smart contract is a script stored in a blockchain network, so it cannot be adulterated and each one has a unique address. This script automatically executes the rules predefined in a contract or consensus, based on these rules it performs transactions among vehicles without the need for an intermediary.

---

[2]Bitcoin: a point-to-point ATM system, Satoshi Nakamoto - https://bitcoin.org/bitcoin.pdf

[3]What is Blockchain, SAP - https://www.sap.com/insights/what-is-blockchain.html

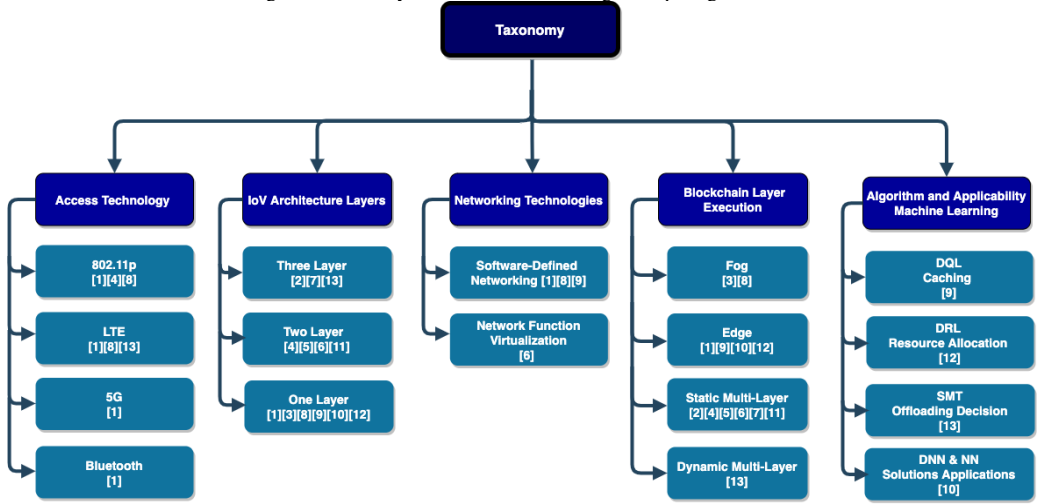

Fig. 2. Taxonomy of Blockchain and Edge computing for IoV

## III. TAXONOMY OF BLOCKCHAIN APPLIED IN IOV

The proposed taxonomy is depicted in Figure 2. The taxonomy contains the building blocks of surveyed architectures that integrate Blockchain and Vehicular Edge Computing for IoV environment. It was structured into five categories: Access network; Layers of the IoV architecture; Networking technologies; Execution layer of blockchain tasks, such as mining, portfolio, consensus, and finally the Algorithms and machine learning techniques applied.

A common goal to add blockchain in an IoV environment is to improve the security of the network. [2] presents a solution for the adequate use of blockchain in vehicular networks with the division of activities and services in a balanced way in the three layers of IoV. The blockchain, in the Perceptual layer, is ensuring network security, the blockchain-related tasks are handled by the Edge-layer, and the Cloud-layer has a backup of the blockchain storage. On the other hand, some articles bring unconventional approaches, like [3]. Aiming to solve the problem of data correctness and the information integrity attacks in a VANET, the study proposes a non-linear blockchain. The blocks are driven by an acyclic graph, where each block has a single transaction.

Considering the paradigm of Electric vehicles cloud and edge (EVCE), [4] uses blockchain in the distributed consensus activities, where they are based on the frequency and contribution of data and energy in the system, which are applied to obtain the score related to a work test (PoW). In [5] consortium blockchain was used, the authors elaborated a scheme utilizing a three-weight subjective logic (TWSL) model to secure storage and sharing of data between vehicle networks.

The paper [6] proposes a many-to-many caching scheme and a trusted access authentication scheme using Blockchain. The authentication scheme is like a mechanism used to achieve real-time monitoring and promote collaborative sharing for vehicles. The blockchain is distributed, being the certifying authority.

The study proposal [7] created a scheme for certificateless authentication, applying the session key independently to each vehicle. They use the blockchain for establishing the V2V groups, sharing, and managing all the existing vehicles of the same group.

Some of the papers utilize software-defined network (SDN) or software-defined vehicular network (SDVN). [1] uses the combination of blockchain and SDN technologies as a way of improving the operation and management of the vehicular networks that employ 5G networks and edge computing. Being the management responsibilities shared between the blockchain and the SDN. To create a virtualized, collaborative, and configurable network, [8] proposes a structure using SDVN along with a paradigm for the blockchain that will allow the certification of transactions to guarantee the anonymity of the data of the nodes in a totally distributed and secure way. In addition, it is proposed a new consensus algorithm called distributed miners connected dominating set (DM-CDS), which dynamically selects miners nodes.

Authors in [9] proposes an optimization-based solution for scaling resources in a virtualized edge computing environment. They use Q-learning technique to determine which server would be the best for a given requested activity. To reach consensus in this network, it was utilized private blockchain.

In the paradigm of connected and autonomous vehicles (CAVs), [10] is seeking to decentralize the ML learning, creating the possibility for vehicles to share the models, learning with each other, and at the same time ensuring privacy and security in the network. To this end, they utilize a blockchain-based collective learning (BCL) technique. The ML algorithms are used to improve the decision-making process of executions activities in the various layers of the IoV network, and the Blockchain is applied to protect the users from security and privacy threats.

The research work [11] is seeking dynamism and security in the exchange of information. So, they present Autonomous Vehicular Social Networks (AVSNs) in the blockchain context. Besides, they feature a consensus algorithm based on reputation, Proof of Reputation (PoR).

In order to alleviate the traffic and reduce latency, [12] proposes to use permissioned blockchain to create a content caching scheme. Due to the high mobility of vehicles, it is a challenge to decide when and where to cache. Therefore, the study seeks to explore a deep reinforcement learning approach to design this scheme. In addition, a new metric for the selection of the block checker, the Utility Proof (PoU), was also proposed.

The research [13] tried to solve the issue of limited computational resources in IoV, proposing a brokerage mechanism to assist the validation decision process whether on-site or in an edge or cloud infrastructure. They used Satisfiability Modulo Theories (SMT) method, besides, the mining of the blockchain can happen anywhere in the network and storage is in every participant node.

The surveyed papers mostly adopted machine learning algorithms as a strategy for decision making, and heuristic for consensus algorithms. It is worth to mention that most papers focused on reinforcement learning and deep learning. Although these techniques were used for different reasons, most researchers have chosen the same techniques when working in an IoV context. It would be interesting to see a comparative study of several decision-making algorithms, to analyze which one is more appropriate to use in an IoV context. It's also worth mentioning that each article used ML for a different purpose. like offloading [13], resource allocation [9], caching [12] or for its application [10].

*A. Scenarios definitions*

We have defined four scenarios to better classify the different architectures of the surveyed papers. Three facets were taken into account to devise these scenarios:
- Which network layer is responsible for the blockchain-related tasks;
- What is the role of the Edge in this scenario;
- What is the resource allocation method in this scenario.

1) Single layer (Fog-layer): In this scenario, the blockchain-related tasks are only in one of the layers, the Fog-layer (In the vehicle's OBU) [3], [8]. The Edge/MEC layer exclusively transmits data between the VANETs and data from the vehicle to the Cloud-layer and vice versa. The Cloud-layer may offer some services to the vehicles, but those would not be blockchain-related. By distributing the blockchain on the final layer, there is less delay between messages exchange, since blockchain-related communications are carried out in V2V mode.

2) Single layer (MEC-layer): In this context, all the blockchain-related tasks are in the Edge/MEC-layer [1], [9], [10], [12]. This layer may offer services that are necessary for the user. The Cloud-layer does not have an important role in this context. As for the Fog-layer, it should only communicate and not do any computation. By moving the computation to the Edge, there is less delay between messages exchanges while maintaining a better processing power than the vehicle's OBU. Both Single layer (Fog-layer) and this scenario have a big issue on resource allocation since all the processing occurs in a single layer.

3) Static multi-layer: In this case, the blockchain-related tasks are also in the Edge/MEC-layer [2], [4]–[7], [11]. Nevertheless, in this scenario, the Fog-layer and the Cloud-layer may handle other tasks, some of those tasks could even be blockchain-related. Each of the layers has its tasks statically assigned, in some way, this may solve the issue of overtaxing the MEC layer, but by failing to analyze the context of the user, sending some of the responsibilities to the cloud may create another set of issues, such as the task never completing due to a congested network. And by never sending some tasks to the cloud, the MEC may be overtaxed in the same way it was in the previous scenario.

4) Dynamic multi-layer: This scenario deals with load-balancing in a hybrid intelligent way [13]. As in the previous scenario, the blockchain-related tasks may be handled in any layer. Nevertheless, in this case, computation resources of each layer and the user's context are taken into consideration to dynamically decide, in which layer each task should be performed. However, if the decision mechanism chooses an inappropriate layer to perform one task, it may cause some of the issues that were mentioned in previous scenarios

*B. Discussion of Blockchain Solutions in IoV*

On Table I we classified:
- The blockchain Type (Public, Private, or Consortium) - It is interesting to highlight that most researches preferred consortium blockchains over other types. Public blockchains usually have low scalability, since they are totally decentralized, every node must process every transaction. There are strategies to overcome this issue, as the architecture proposed on [3]. On Public blockchains there's also the issue with storage since every node must store every block of the immutable blockchain. Private blockchains have their appeal since it does not have most of the issues a public blockchain has. A central authority decides which nodes can enter the blockchain, which nodes can be miners, which nodes must have all transactions, etc. But it's more centralized since a single company manages the whole blockchain. A consortium blockchain, is only partially private since several companies share the ownership of the blockchain
- The blockchain Layer, i.e. where the mining and the consensus are mostly happening. This column is relevant for the reader to know which layers were used in a static or dynamic multi-layer architecture.
- Architectural Approach, using the previously defined architecture models. Bringing the blockchain closer to the

TABLE I
SUMMARY OF SOLUTIONS: BLOCKCHAIN FEATURES.

| Paper | BlockchainType | BlockchainLayer | ConsensusAlgorithm | Architectural Approach |
|---|---|---|---|---|
| Jianbin Gao 2019 [1] | Consortium | Edge | PBFT | Edge-layer |
| Shaoyong Guo 2019 [6] | - | Edge&Cloud | - | Static multi-layer |
| Jiawen Kang 2019 [5] | Consortium | Fog&Edge | PoW | Static multi-layer |
| Yueyue Dai 2019 [12] | private/permissioned | Edge | PoU | Edge-layer |
| Heena Rathore 2019 [3] | - | Fog | TangleCV Policy | Fog-layer |
| Youcef Yahiatene 2018 [8] | Public&Private | Fog | Authors created DM-CDS | Fog-layer |
| XiaoDong Zhang 2018 [2] | - | Fog&Edge&Cloud | - | Static multi-layer |
| Chau Qio 2018 [9] | private/permissioned | Edge | PBFT | Edge-layer |
| Haowen Tan 2019 [7] | Consortium | Fog&Edge&Cloud | - | Static multi-layer |
| Vincenzo 2019 [13] | - | Fog&Edge&Cloud | PoS+PoW | Dynamic multi-layer |
| Hong Liu 2018 [4] | Consortium | Fog&Edge | PoEC/PoW + PoF/PoS | Static multi-layer |
| Yuchuan Fu 2019 [10] | Consortium | Edge | BFT-DPoS | Edge-layer |
| Y.Wang 2020 [11] | Consortium | Fog&Edge | PoR | Static multi-layer |

users decreased latency and increased the overall quality of service since the information got to the user faster [3], [8]. But not every vehicle is so well equipped that can do all storage and computation. Some researchers decided to delegate these blockchain tasks to the Edge or cloud layer. In fact, most papers have chosen to leave the blockchain in the edge or create a hybrid Fog-Edge or Fog-Edge-Cloud architecture. This was done to avoid overtaxing the vehicles with processing and storage

- Consensus Algorithm - Here we specify which consensus algorithm was used in each surveyed paper.

  1) PoW - Proof-of-work, one of the most known consensus algorithms, used by BitCoin. Every node competes with one another to solve a challenge, the nodes with the best computational power usually win the competition and generate the new block. This makes it harder for attackers since the cost of attacking the chain, trying to pass a new block with fraudulent transactions is too high. But that puts a strain in the mining nodes since for every consensus there is a need for competition. In a VANET environment, if a researcher chooses this method, then it would be best to leave the mining outside the vehicles. [5], and [13] both used PoW, but combined with another strategy, so the difficulty of the challenge would be smaller for certain individuals. This eases the energy consumption burden since the challenge would be solved faster. [11]'s PoR also uses PoW competition in the end. [4] also used a combination of algorithms, PoW and Proof-of-Storage, the nodes that are using the most storage gain extra coins each consensus. But the authors didn't change PoW difficulty and its disadvantages remain the same.

  2) PoS - Proof-of-Stake, also a very common consensus algorithm used by NXT cryptocurrency. This algorithm came to try to solve PoW inherent issue of energy consumption. This approach chooses the miner based on how many coins the node has accumulated. The main issue is the lack of variation randomization. Usually, rich nodes would always win the miner election, and that may be a security issue. This algorithm was only present in the surveyed papers if it was combined with another, as it was the case for [5] and [13].

  3) PoU - Proof-of-utility, quite similar to DPoS, in this algorithm we run an election, where the nodes vote for which mining node has the higher utility. This utility can be calculated using multiple variables that change their weight over time, making these variables non-linear distributed. These variables could include stake, participation rates, seniority, this can vary as it depends on the implementation. PoU is scalable and safe. [12] authors calculated this Utility based on the amount of time the node takes to provide information for the vehicle. The main issue with this approach is how the Utility is calculated. Since the number of variables and their weights is all open for the developer to choose, a bad choice could cause a major security issue. So, it's hard to implement it safely and correctly.

  4) PoR - Proof-of-Reputation. The PoR devised in [11] is really similar to PoU or DPoS. There is an election, and the node reputation is directly proportional to its voting power. The main difference is that, after the election, the elected nodes won't mine, they will compete in a PoW challenge, and only the winner is going to be rewarded. The higher their reputation, the smaller the difficulty of the challenge is. This combination of DPoS and PoW suppress most of PoW shortcomings, since we don't need all the network to compete, and those who do compete don't expend that much energy since the difficulty decreases as the reputation increases. But there's a risk of centralization

  5) PBFT - Proof Bizantine Fault Tolerance is a non-anonymous, message heavy algorithm. If 7 nodes are trying to reach consensus 71 messages are necessary. If a node is considered in the traditional sense, as a RSU or a vehicle, then it doesn't scale.

It is possible, however, in a consortium blockchain where each organization elects a single node to participate in the consensus. The paper [1] does not mention this approach.

6) BFT-DPos - Bizantine fault tolerance delegated proof-of-stake. This algorithm puts together two concepts; the Delegated proof-of-stake is used to choose the nodes that will try to achieve consensus-based on their stake. After choosing 21 nodes, these nodes will continuously produce 12 blocks. And for each one they apply the second concept, the BFT to achieve consensus, verifying the blocks. This is a scalable and fast way to generate blocks. Since the users have a tiny chance to influence the result, in a real-life scenario, is likely that the users would deposit their money on an exchange, so the exchange can vote for them. That can cause voting centralization, this issue is commonly seen in DPoS.

7) TangleCV - To understand the tangleCV policy, it's necessary to explain the tangle architecture. It's based on blockchain, but it takes a different approach. The blocks might be chained in an acyclic way, each block has only one transaction and each new transaction requires PoW validation of two previous transactions. After validation, a transaction gains a weight calculated based on the amount of necessary work to validate that transaction, and this weight is its trustworthiness. The tangleCV Policy states that a transaction that has a weight higher than a threshold might self-validate, otherwise it has to do a PKI authentication to validate. In the tangle architecture, this policy has many advantages such as reducing bandwidth congestion and removing delays. But there was no security test done in [3]. Also, it's difficult to compare this policy with other consensus algorithms, since the architecture is a little different.

8) DM-CDS - The Distributed miners connected dominating set algorithm is based on another algorithm, DSP-CDS or Distributed single-phase connected dominating set. DSP-CDS is used to decide if a given node should enter a set of dominating nodes, it adapts well to changes in the network topology. DM-CDS uses a miner score function that considers several parameters, such as a trust model, degree of connectivity, average link quality, to decide which node may validate the transactions. This algorithm is a little similar to PoS where instead of stake, the main factor that decides is the miner score. And since the miner score is variable with time, the miners would variate more than in PoS.

## IV. Conclusion

The Internet of Vehicles (IoV) will play an important role in future smart cities, providing road safety, convenience and several other benefits. Therefore, it is necessary to search for the guarantee of safety in the use of integrated technologies of ITS. This study sought to identify the types of blockchain, consensus algorithms, as well as their uses, contexts, and operation in the environment (cloud, edge, fog, and hybrid). We sought to categorize the different usage scenarios, taking into account their various positive and restrictive aspects of each proposal, be it communication, assistance, security, application, consensus, and intelligent decision-making technologies, all of these aspects are present in Figure 2, then we classified each paper into one of these scenarios. We discussed each of the proposals of Table 1, delivering a comparative study of how blockchain and edge computing are being used in IoV, how they are being designed, which algorithms and technologies were used, listing their advantages and disadvantages. In summary, this survey aimed to guide and assist researchers, analysts, developers in their decision making of current solutions or future proposals for the IoV context. In the future we will carry out an analysis on delay and energy efficiency of those consensus algorithms in an unified simulation framework, so they can be properly compared.